\documentclass[aps, pra, reprint, superscriptaddress, showpacs, amsmath, amssymb]{revtex4-1}
\usepackage{txfonts}
\usepackage[pdftex]{graphicx}
\usepackage{dcolumn}
\usepackage{bm}
\usepackage{soul}
\usepackage{color}
\usepackage{hyperref}
\hypersetup{
    colorlinks=true,        
    linkcolor=blue,          
    citecolor=blue,        
    filecolor=magenta,      
    urlcolor=blue           
}

\begin{document}

\title{Spin state ordering of strongly correlating LaCoO$_{3}$ induced at ultrahigh magnetic fields}
\author{Akihiko~Ikeda}
\email[Corresponding author: ]{ikeda@issp.u-tokyo.ac.jp}
\author{Toshihiro~Nomura}
\author{Yasuhiro~H.~Matsuda}
\email[Corresponding author: ]{ymatsuda@issp.u-tokyo.ac.jp}
\author{Akira~Matsuo}
\author{Koichi~Kindo}
\affiliation{Institute for Solid State Physics, University of Tokyo, Kashiwa, Chiba, Japan}
\author{Keisuke~Sato}
\affiliation{Ibaraki National College of Technology, Department of Natural Science, Hitachinaka, Ibaraki, Japan}

\date{\today}

\begin{abstract}
Magnetization measurements of LaCoO$_{3}$ have been carried out up to 133 T generated with a destructive pulse magnet at a wide temperature range from 2 to 120 K.
A novel magnetic transition was found at $B>100$ T and $T>T^{*}=32\pm 5$ K which is characterized by its transition field increasing with increasing temperature.
At $T<T^{*}$, the previously reported transition at $B\sim65$ T was observed. Based on the obtained $B$-$T$ phase diagram and the Clausius-Clapeyron relation, the entropy of the high-field phase at 80 K is found to be smaller for about $1.5$ J K$^{-1}$ mol$^{-1}$ than that of the low-field phase.
We suggest that the observed two high-field phases may originate in different spatial orders of the spin states and possibly other degrees of freedom such as orbitals.
An inherent strong correlation of spin states among cobalt sites should have triggered the emergence of the ordered phases in LaCoO$_{3}$ at high magnetic fields.
\end{abstract}


\maketitle

Due to the strong correlations between the electrons, the transition metal oxide serves as a vast field hosting rich electronic phases represented by high-temperature superconductivity, colossal magnetoresistance and magnetic-field-induced ferroelectorics \cite{Imada, Tokura}.
Among them, cobalt oxides are unique for their spin state degrees of freedom which not only bring about a magnetic crossover but also a metal-insulator transition (MIT) \cite{Tachibana} in the thermal evolution.
Perovskite cobalt oxide, LaCoO$_{3}$, has attracted significant attention for more than five decades for its unusual magnetic and transport properties, namely, the crossover from a diamagnet to a Curie paramagnet at 100 K and the transition from a paramagnetic insulator to a paramagnetic metal at 500 K with increasing temperature \cite{Goodenough1958}.
Within the ionic picture, possible spin states of Co$^{3+}$ are the low spin state (LS: $t_{2g}^{6}e_{g}^{0}$, $S=0$) and the high spin state (HS: $t_{2g}^{4}e_{g}^{2}$, $S=2$) that energetically lie close to each other due to the delicate balance of Hund's coupling and crystal field splitting.
Besides those, the intermediate spin state  (IS: $t_{2g}^{5}e_{g}^{1}$, $S=1$) is also argued to be stabilized due to the strong hybridization with the O $2p$ state \cite{Korotin}.
Representative ideas describing the spin states of LaCoO$_{3}$ in the temperature range above 100 K are (i) the LS-HS mixture state \cite{Raccah, Haverkort, Knizek, Ropka2003, Kyomen2003, Kyomen2005} and (ii) the IS state \cite{Korotin, Yamaguchi1997, Ishikawa}.
However, they are still controversial.
It is notable that recent theoretical studies on the two-orbital Hubbard model have qualitatively reproduced the thermally induced spin crossover and MIT with paramagnetic local moments \cite{Kunes2011, Kanamori2011, Krapek}.
On the other hand, they are inclined to predict the ordering of different spin states which is not found experimentally except for a few studies \cite{Doi}.

The validity of the models on spin states should be well judged by their field effects.
One can uncover magnetic excited states using high magnetic fields at low temperatures, eliminating the thermal effect. 
Thermodynamical properties of the magnetic phase can also be revealed by observing its temperature and magnetic field dependence \cite{Tokunaga, Nomura}.
In the case of LaCoO$_{3}$, a spin gap of about 100 K \cite{Yamaguchi1996} necessitates a high magnetic field amounting to 100 T.
In fact, a first-order field-induced spin state transition \cite{Sato2009, Moaz} accompanied by magnetostriction \cite{Moaz, Rotter} has been found at $B=65$ and 70 T with magnetization measurements up to 100 T at below 4.2 K.
The results are either understood in terms of the local spin crossover model \cite{Sato2009} or the formation of the spin state crystalline (SSC) phase, where the different spin states at Co$^{3+}$ and possibly the orbitals are spatially ordered \cite{Moaz, Rotter} and further, the following two magnetization jumps at $B>100$~T are predicted by the Ising type SSC model \cite{Moaz}.
With the explosive magnetic flux compression technique, magnetization up to 3.5$\mu_{\mathrm{B}}$ was observed at 500 T, 4.2 K, although the smeared transitions up to 100~T may be due to the fast sweeping rate ($>10$~T/$\mu$s) \cite{Platonov}.
The $B$-$T$ range explored so far, however, has been limited to low temperatures.
To verify the physical origins of the thermally induced magnetic phase and the field-induced magnetic phase of LaCoO$_{3}$, it is plausible to explore the properties of LaCoO$_{3}$ in even wider $B$ and $T$ ranges and clarify how those phases evolve and interact with each other on the $B$-$T$ plane.

In this Rapid Communication, we report a high-field magnetization study of LaCoO$_{3}$ up to 133 T at a wide range of temperatures from 2 to 120 K, from whose data a phase diagram in the wide $B$-$T$ range is constructed.
We found first-order magnetic transitions at $B>100$ T at $T>T^{*}=32\pm 5$ K where the transition fields increased with increasing $T$, suggesting the existence of the low entropy phase at $B>100$ T and at $T>T^{*}$.
We also confirmed the reported first order magnetic transition at $\sim65$ T at $T<T^{*}$ where the transition field was almost temperature independent.
We obtained a rich phase diagram that contradicts the prediction based on the spin crossover in the local ion picture.
We discuss the result in light of the formation of the field induced ordered phase due to strongly correlating spin states and other degrees of freedom such as orbitals.

\begin{figure*}[t]
\begin{center}
\includegraphics[angle=0, scale=0.7, clip]{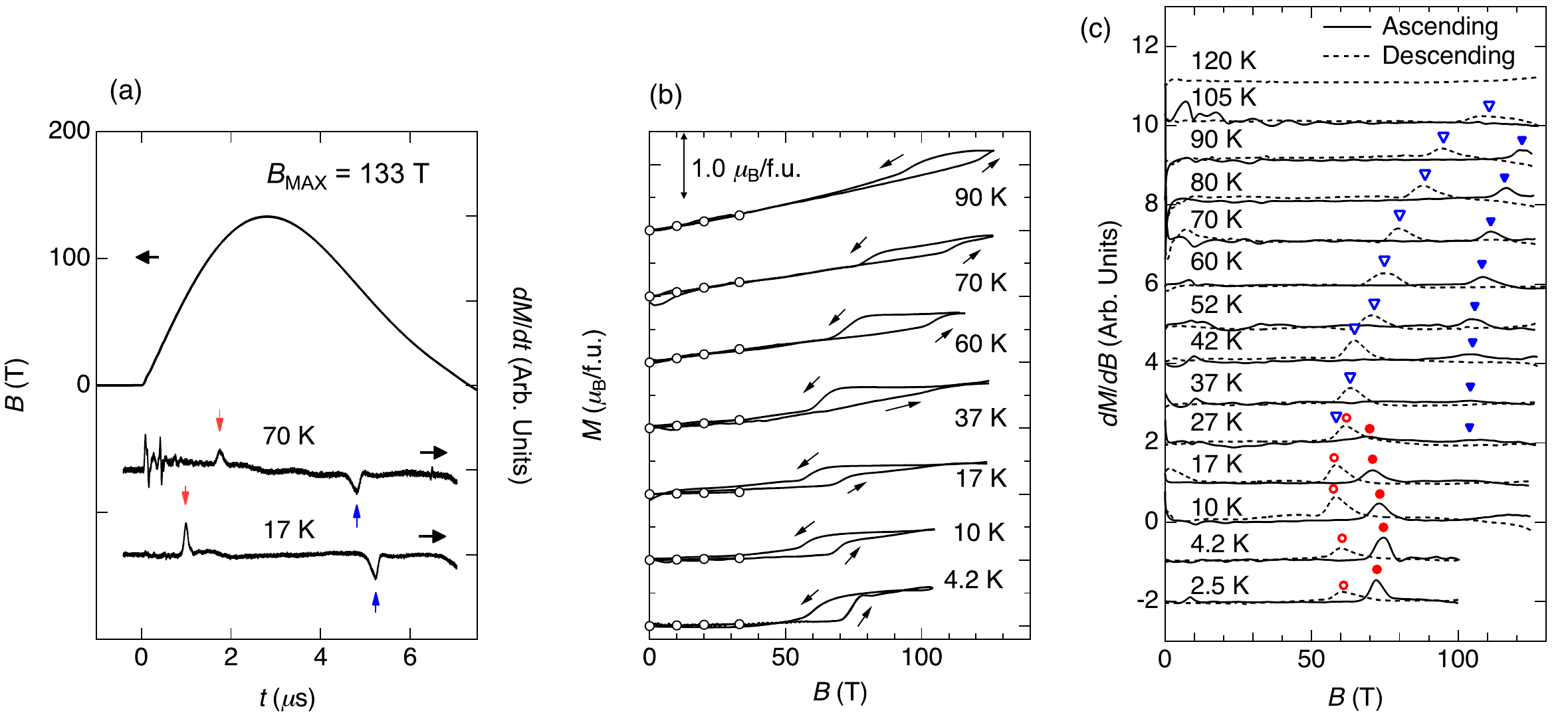} 
\caption{(a) Time evolution of the magnetic field and the time derivative of magnetization $dM/dt$ which is proportional to the induction voltage of LaCoO$_{3}$ at 17 and 70 K. The arrows pointing downwards and upwards denote the peaks in $dM/dt$ curves. (b) Magnetization $M$ of LaCoO$_{3}$ as a function of magnetic field at various temperatures obtained using the single turn coil (thin curves). The data up to 33 T reported by Hoch \textit{et al.} using static fields (open circles) were adopted from Ref. \onlinecite{Hoch}. (c) Magnetic field derivative of the magnetization $dM/dB$ curves of LaCoO$_{3}$ as a function of $B$ obtained using the single-turn coil in the ascending field (solid curves) and the descending field (dashed curves). The solid and open symbols denote the magnetic transition in the ascending and the descending fields, respectively.  \label{fig1}}
\end{center}
\end{figure*}

High field magnetization measurements were carried out in the following manner. For the generation of a high field with a maximum field $B_{\mathrm{Max}}$ of 133 T, a horizontal type single-turn coil, a semi-destructive pulse magnet \cite{Miura}, was employed. Helium flow type cryostats made of nonmetallic parts were used to cool the sample \cite{Takeyama1988, Amaya}. The temperature at the sample space ranged from 10 to 120 K and was measured with a chromel-constantan thermocouple. We also used a vertical type single turn coil with a helium bath type cryostat for the measurements at $T=2.5$ and 4.2 K and $B_{\mathrm{Max}}\sim105$ T, as described in Ref. \cite{Takeyama2012}. The magnetization ($M$) of LaCoO$_{3}$ was obtained by measuring the induction voltage (proportional to $dM/dt$) of a well compensated pair of pickup coils, one of which held the sample inside. Small grains of single crystalline LaCoO$_{3}$ \cite{Sato2009} were put into a sample space of $\phi=0.9$ and $l=3$ mm with their crystal axis unoriented. The magnetic field $B$ was measured with a calibrated pickup coil placed close to the sample space. 

Representative results of the time derivative of $M$, $dM/dt$, at 17 and 70 K are shown in Fig.  \ref{fig1}(a), along with the time evolution of $B$. At 17 K, sharp peaks are seen at 70 and 60 T, respectively, whereas, at 70 K, the peaks were observed at higher fields, indicating that the transition fields are temperature dependent. $M$ curves were obtained by numerically integrating the $dM/dt$ data. $dM/dB$ curves were obtained by dividing the $dM/dt$ data with the $dB/dt$ data. They are plotted against $B$, as shown in Figs. \ref{fig1}(b) and \ref{fig1}(c), respectively. In Fig. \ref{fig1}(b), absolute values are evaluated by scaling the data to the $M$ data obtained using a nondestructive pulse magnet at ISSP, Univerisity of Tokyo. The obtained $M$ curves in Fig. \ref{fig1}(b) are in good agreement with the low-field magnetization data up to 33 T, as reported in Ref. [\onlinecite{Hoch}]. Whereas the magnetic transitions at $\sim65$ T and at below 30 K have been reported previously \cite{Sato2009, Moaz, Rotter}, we show the magnetic transitions at above 37 K and above 100 T.  One can clearly notice that the field-induced magnetic transitions in Fig. \ref{fig1}(b) and \ref{fig1}(c) are temperature dependent. This trend may have a common root with the observed positive temperature dependence of magnetic transitions at $B\simeq$ 60 T and at $T$ > 40 K in Ref. \cite{Moaz, Rotter}.

\begin{figure}[t]
\begin{center}
\includegraphics[angle=0, scale=0.43, clip]{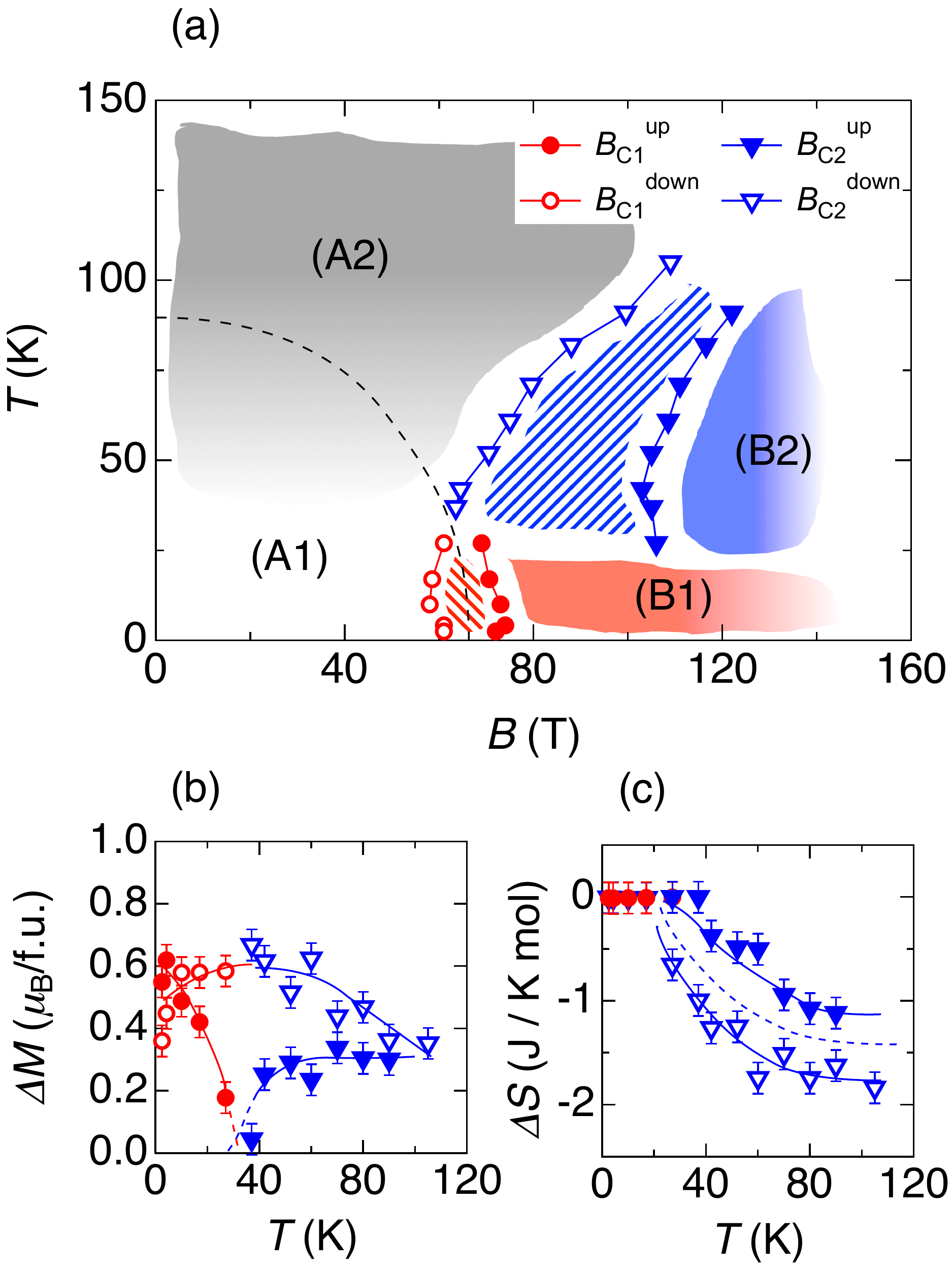}
\caption{(a) The $B$-$T$ phase diagram of LaCoO$_{3}$ based on the observed transition fields $B_{\mathrm{C1}}$ and $B_{\mathrm{C2}}$ in the present study. The dashed curve represents the predicted phase boundary based on the spincross over in the local ion picture \cite{Biernacki2005}. (b) $\Delta M$ and (c) $\Delta S$ at the phase boundary at each temperature. The same symbols are used as in (a). \label{fig2}}
\end{center}
\end{figure}

We first focus on the obtained $M$ at 4.2 K in Fig.  \ref{fig1}(b). The amount of the magnetization jump $\Delta M$ at the transition is $\sim 0.5 \mu_{\mathrm{B}}/\mathrm{f.u.}$, which is in good agreement with the reported values \cite{Sato2009, Moaz}. The values of the transition field observed in the ascending and descending fields are $\sim75$ T and 65-60 T, respectively. The existence of a large hysteresis of about 15 T indicates that the transition is a first-order transition. The relatively smeared transition in the descending fields in Fig. \ref{fig1}(b) should originate from the heating effects during the first order phase transition, as suggested in Ref. \cite{Moaz}. With increasing temperature up to 27 K, the transitions become smeared, possibly due to the thermal effect, as seen in Fig. \ref{fig1}(c). Our result is inconsistent with the reports in Ref. [\onlinecite{Moaz}], where the first increase of $M$ of $\sim 0.5 \mu_{\mathrm{B}}/\mathrm{f.u.}$ at 63 T was followed by a second increase of $M$ of $\sim 0.5 \mu_{\mathrm{B}}/\mathrm{f.u.}$ at $\sim70$ T. The cause of the discrepancy is not clear at this moment, although it may be due to the sample or the field sweeping rate dependence.  We regard that the second transition is absent in the present study. The sweep rate up to 100 T is faster in our case ($\sim$50 T/$\mu$s) than in the case of the explosive compression technique ($\sim$10 T/$\mu$s) \cite{Platonov}. Therefore, the smearing of the sharp transition in Ref. \cite{Platonov} may not be due to the intrinsic effects, such as thermal effects.

Next, we observe the temperature dependence of $M$ and $dM/dB$ in Figs.  \ref{fig1}(b) and \ref{fig1}(c), respectively. Guided by the sudden change in the transition field at $T^{*}=32\pm5$ K, we term the transition fields for the ascending field and descending field at $T<T^{*}$ as $B^{\mathrm{up}}_{\mathrm{C1}}$ and $B^{\mathrm{down}}_{\mathrm{C1}}$ denoted with solid and open circles, and at $T>T^{*}$, $B^{\mathrm{up}}_{\mathrm{C2}}$ and $B^{\mathrm{down}}_{\mathrm{C2}}$ denoted with solid and open triangles, respectively. At $T>T^{*}$, we found that the novel magnetic transition is present at $B > 100$ T in the ascending field ($B^{\mathrm{up}}_{\mathrm{C2}}$) as denoted by the solid triangles in Fig.  \ref{fig1}(c). With increasing temperature, the peaks at $B^{\mathrm{up}}_{\mathrm{C2}}$ in Fig.  \ref{fig1}(c) are gradually sharpened and shifted towards higher fields. $B^{\mathrm{down}}_{\mathrm{C2}}$ also shifted to higher fields with increasing temperature at $T>T^{*}$. This is highly in contrast with $B^{\mathrm{up}}_{\mathrm{C1}}$ and $B^{\mathrm{down}}_{\mathrm{C1}}$ at $T<T^{*}$ being independent of temperature \cite{Sato2009, Moaz}.

We plot the obtained transition fields on the $B$-$T$ plane as shown in Fig. \ref{fig2}(a). The transitions at $T>T^{*}$ and $B>100$ T are reported (colored in blue). The hysteresis region is indicated by the shaded area. For clarity, we term the low-temperature low-field region and the high-temperature low-field region to be phases (A1) and (A2), respectively. We also term the high-field phases (B1) and (B2). It is evident that the high 
-field phases (B1) and (B2) are separated from the low field phases (A1) and (A2) by a first-order magnetic transition with hysteresis. Phases (B1) and (B2) are distinguished based on $T^{*}$.

By integrating the peaks in the $dM/dB$ curves, we obtained $\Delta M$ at each transition field for various temperatures, as shown in Fig. \ref{fig2}(b). The saturation magnetizations $M_{\mathrm{S}}$ expected for IS or HS Co$^{3+}$ are 2.0$\mu_{\mathrm{B}}/\mathrm{f.u.}$ or 4.0$\mu_{\mathrm{B}}/\mathrm{f.u.}$, respectively, provided $g=2$. $M_{\mathrm{S}}$ is not reached even after the magnetic transition at 70 K ($M\sim 1.0 \mu_{\mathrm{B}}/\mathrm{f.u.}$). With the observed values of $\Delta M$ in Fig. \ref{fig2}(b) and $dB/dT$ obtained from the phase boundary in Fig. \ref{fig2}(a), we deduced the entropy change $\Delta S$ at the field-induced transition based on the Clausius-Clapeyron relation \cite{Landau} $dB/dT=-\Delta S/\Delta M$, as shown in Fig. \ref{fig2}(c).  At $T<T^{*}$, the slope is vertical resulting in $\Delta S \sim 0$~J~K$^{-1}$ mol$^{-1}$. At $T>T^{*}$, $\Delta S$ gradually decreases from 0 and converges to $\sim -1.5$ J K$^{-1}$ mol$^{-1}$ at $T > 80$~K. This compares to the entropy increase at the thermally induced spin crossover in LaCoO$_{3}$ of $\sim 2.0$~J~K$^{-1}$~mol$^{-1}$ from 13 to 80 K \cite{stolen}. Both of them are much smaller than the value of $R \ln 3=$ 9.13 J K$^{-1}$ mol$^{-1}$ expected for the thermal spin crossover from $S=0$ to $S=1$.

The magnetic transitions discovered at $B>100$~T and $T>T^{*}$ in the present study need temperature assistance to ascertain their origin. Our data set lacks the low-temperature ($T < 4.2$~K) high field ($B > 100$~T) data, which may make the absence of a magnetic transition at $100$ T $<B<140$~T and at $T<T^{*}$ inconclusive. However, the data at 4.2 K up to 140 T are actually provided in Ref. \cite{Moaz} by making use of the single-turn coil, evidencing that such a magnetic transition is absent up to 140 T. On this basis, we regard the magnetic transitions discovered at $B>100$~T and $T>T^{*}$ to have different origins from the predicted spin state cascade based on the Ising type SSC model at 0 K in Ref. \cite{Moaz}, where the predicted magnetic transition at $B>100$~T should be present even at $T\sim$ 4.2 K, and it is not predicted that another ordered phase such as (B2) appears with increasing temperature.


Here, we argue that the most striking feature of the obtained phase diagram shown in Fig. \ref{fig2}(a) is that the transition field increases with increasing temperature. It is completely contrary to the shared tendency of the previous reports on spin crossover compounds such as cobalt oxides [Sr$_{1-x}$Y$_{x}$CoO$_{3}$ \cite{Kimura2008}, (Pr$_{1-y}$Y$_{y}$)$_{0.7}$Ca$_{0.3}$CoO$_{3}$ \cite{Marysko, Naito2014}] and coordinate compounds (Fe[(phen)$_{2}$(NCS)$_{2}$] \cite{Qi}, [Mn$^{\mathrm{III}}$(taa)] \cite{Kimura2005}), where the transition fields are observed to decrease with increasing temperature, as schematically shown by the dashed curve in Fig. \ref{fig2}(a). This tendency can be readily anticipated by considering the spin crossover in the local ion picture, where the ground state is less magnetic (i.e., LS) and that the excited state is more magnetic (e.g., HS or IS). In this situation, the magnetic state will be occupied with increasing either $T$ or $B$ due to the entropy or Zeeman energy contribution, respectively. In LaCoO$_{3}$, the ground state is the LS phase \cite{Asai1994}, denoted as phase (A1) in Fig. \ref{fig2}(a), whose entropy is considered to be small. The thermally induced paramagnetic state \cite{Asai1994}, denoted as phase (A2) in Fig. \ref{fig2}(a), is considered to possess a larger entropy due to the magnetic, orbital, and phonon degrees of freedom of HS or IS species and the mixing entropy of the LS-HS or LS-IS complexes \cite{Biernacki2005}. Based on the local model for spin crossover, it is expected that the transition field decreases with increasing temperature and that phase (B1), (B2) merges with phase (A2) at the high-temperature and high-field region, as shown by the dashed curve in Fig. \ref{fig2}(a). It is clear that phase (A2) and phase (B1), (B2) are the distinct phases in the present result as shown in Fig. \ref{fig2}(a). It is now decisive that the local model for the spin crossover compounds \cite{Kyomen2003, Biernacki2005} is not applicable to the $B$-$T$ phase diagram of LaCoO$_{3}$, suggesting that phasesr (A2) and (B1), (B2) are distinct in origin, which is contrary to the previous notion \cite{Sato2009, Rotter}.

We now discuss the origin of the observed high-field phases (B1) and (B2). In the present observation, the reduction of $S$ is observed in the transition from phase (A2) to phase (B2), as shown in Fig. \ref{fig2}(c). This may suggest that some order is present in phase (B2). The candidates for the order of phase (B2) are (i) antiferromagnetic order (AFM), (ii) spin state crystalline (SSC), and (iii) orbital order (OO). In the SSC, the spin states of Co$^{3+}$ are spatially ordered. Among them, we believe the SSC is the most plausible idea for the following reasons. First, because AFM becomes unstable under larger magnetic fields, its N\'{e}el temperature is expected to decrease with increasing magnetic field. However, as shown in Fig. \ref{fig2}(a), the transition temperature of (B2) increases with increasing magnetic field. Therefore, AFM is excluded. Next, we consider the SSC. In phase (A2) the spin states are disordered. At the magnetic transition, the number of Co$^{3+}$ in the magnetic spin states is increased and the spatial order of the spin states is obtained, forming the SSC, the spatial order of spin states. This scenario is in good agreement with experimental observations, namely, the sudden increase of magnetization and the decrease of entropy. Thus, we regard the SSC is present in phase (B2). Lastly, we consider OO. The orbital degree of freedom is quite spin state dependent. Therefore, if the spin states are disordered, it should be very difficult for the OO to appear. Besides, OO itself does not change the magnetization. Therefore, OO alone cannot be the order parameter of phase (B2). On the other hand, the OO on the background of the SSC should appear plausible. Such spin state ordering is also suggested in recent theories \cite{Knizek, Kunes2011, Kanamori2011, Krapek} and high-field experiments \cite{Moaz, Rotter}.

Another feature found in the obtained phase diagram in Fig. \ref{fig2}(a) is the sudden change in the transition fields at $T^{*}$, making the two high-field phases (B1) and (B2) distinct. The phase boundary between phase (B1) and (B2) seems horizontal ($dT/dB=0$) at $T^{*}$. This means that $\Delta M/\Delta S=0$ in the virtual transition from phase (B1) to (B2) based on the Clausius-Clapeyron relation. We deduce $\Delta M=0$, assuming that $\Delta S$ is not so large. As a possible origin of the two distinctive phases (B1) and (B2), we discuss that, besides SSC, another order may be present in phase (B1) which does not change $M$. This is because the SSC of phase (B2) is expected to be even more stable in phase (B1) due to the lower temperature.

Possible origins for the order of (B1) in addition to the SSC are (i) AFM, (ii) OO, (iii) excitonic condensate (EC), and (iv) the SSC with a spatial pattern that is different from (B2). We note here that it is difficult at present to further qualify those possibilities, except for the AFM. The AFM in phase (B1) is excluded because $M$ should be smaller than that of phase (B2). This is in contradiction to the experimental observation. EC may be plausible, although further experimental evidence is needed to confirm it. EC has been recently proposed as the origin of the insulating phase of LaCoO$_{3}$ and Pr$_{0.5}$Ca$_{0.5}$CoO$_{3}$ \cite{Kunes2014, Kunes001, Kunes002, Kaneko2012, Kaneko2014, Kaneko2015}. In a very recent report, it is predicted by a dynamical mean field model calculation that a field-induced EC is possible \cite{Sotnikov}.  Switching between two different SSCs may also be possible. The SSCs with various spatial patterns were considered with generalized gradient approximation (GGA+$U$) calculations in Ref. [\onlinecite{Knizek}]. Some two SSCs with the same $M$ may undergo a temperature-induced transition from one to the other with the assistance of the entropy difference of those phases due to a lattice or orbital contribution.

The OO in phase (B1) is also in good agreement with the experimental results. Co$^{3+}$ in the IS or HS both possess orbital degrees of freedom at the $e_{g}$ and $t_{2g}$ orbitals, respectively. The formation of the OO at phase (B1) will stabilize it energetically, which may well result in the reduction of the transition field to phase (B1) as compared to that to phase (B2), being in accord with the observed change of the transition field at $T^{*}$. In addition, the flat phase boundary between (B1) and (B2) is also in good agreement with the order-disorder phase transition of orbitals \cite{Murakami} or the switching between different OO \cite{Mcqueen} because they can occur with $\Delta M=0$. In those cases, orbitals are ordered in phase (B1) and in (B2) the orbitals are disordered or forming the OO with different spatial pattern. For these reasons, we regard that, in phase (B1), OO may be present in addition to the SSC. Orbital ordering taking place along with the spin state ordering has also been claimed in YBaCo$_{2}$O$_{5}$ \cite{Vogt}, Sr$_{3}$YCo$_{4}$O$_{10.5}$ \cite{Nakao}, the thin film of LaCoO$_{3}$ \cite{Fujioka}, and a previous high-field study on LaCoO$_{3}$ \cite{Moaz}. We note, however, the origin of phases (B1) and (B2) is still an open question to be explored in future studies.

In conclusion, high-field magnetization measurements of LaCoO$_{3}$ up to 133 T were carried out in a wide temperature range from 2 to 120 K. At $T>T^{*}$, we observed the novel magnetic transition at $B>100$ T. In addition, we observed the previously reported magnetic transition at $\sim65$ T with $T<T^{*}$. Based on the obtained $B$-$T$ phase diagram and the Clausius-Clapeyron relation, it was found that the high-field phases possess lower entropy than the low-filed phases, and that the high-field phases are separated into two phases at $T=T^{*}$. We argue that the observed magnetic transitions take place from the LS-HS or LS-IS disordered phase to the ordered SSC of LS-HS or LS-IS complex. At $T<T^{*}$, spatially different SSC or orbital order may develop.

The authors acknowledge M. Tokunaga and S. Ishihara for fruitful discussions, T. T. Terashima, S. Takeyama and K. Yoshikawa for experimental and various supports. This work was supported by JSPS KAKENHI Grant-in-Aid for Young Scientists (B) Grant Number 16K17738.

\bibliography{lco}

\begin{thebibliography}{50}%
\makeatletter
\providecommand \@ifxundefined [1]{%
 \@ifx{#1\undefined}
}%
\providecommand \@ifnum [1]{%
 \ifnum #1\expandafter \@firstoftwo
 \else \expandafter \@secondoftwo
 \fi
}%
\providecommand \@ifx [1]{%
 \ifx #1\expandafter \@firstoftwo
 \else \expandafter \@secondoftwo
 \fi
}%
\providecommand \natexlab [1]{#1}%
\providecommand \enquote  [1]{``#1''}%
\providecommand \bibnamefont  [1]{#1}%
\providecommand \bibfnamefont [1]{#1}%
\providecommand \citenamefont [1]{#1}%
\providecommand \href@noop [0]{\@secondoftwo}%
\providecommand \href [0]{\begingroup \@sanitize@url \@href}%
\providecommand \@href[1]{\@@startlink{#1}\@@href}%
\providecommand \@@href[1]{\endgroup#1\@@endlink}%
\providecommand \@sanitize@url [0]{\catcode `\\12\catcode `\$12\catcode
  `\&12\catcode `\#12\catcode `\^12\catcode `\_12\catcode `\%12\relax}%
\providecommand \@@startlink[1]{}%
\providecommand \@@endlink[0]{}%
\providecommand \url  [0]{\begingroup\@sanitize@url \@url }%
\providecommand \@url [1]{\endgroup\@href {#1}{\urlprefix }}%
\providecommand \urlprefix  [0]{URL }%
\providecommand \Eprint [0]{\href }%
\providecommand \doibase [0]{http://dx.doi.org/}%
\providecommand \selectlanguage [0]{\@gobble}%
\providecommand \bibinfo  [0]{\@secondoftwo}%
\providecommand \bibfield  [0]{\@secondoftwo}%
\providecommand \translation [1]{[#1]}%
\providecommand \BibitemOpen [0]{}%
\providecommand \bibitemStop [0]{}%
\providecommand \bibitemNoStop [0]{.\EOS\space}%
\providecommand \EOS [0]{\spacefactor3000\relax}%
\providecommand \BibitemShut  [1]{\csname bibitem#1\endcsname}%
\let\auto@bib@innerbib\@empty
\bibitem [{\citenamefont {Imada}\ \emph {et~al.}(1998)\citenamefont {Imada},
  \citenamefont {Fujimori},\ and\ \citenamefont {Tokura}}]{Imada}%
  \BibitemOpen
  \bibfield  {author} {\bibinfo {author} {\bibfnamefont {M.}~\bibnamefont
  {Imada}}, \bibinfo {author} {\bibfnamefont {A.}~\bibnamefont {Fujimori}}, \
  and\ \bibinfo {author} {\bibfnamefont {Y.}~\bibnamefont {Tokura}},\ }\href
  {\doibase 10.1103/RevModPhys.70.1039} {\bibfield  {journal} {\bibinfo
  {journal} {Rev. Mod. Phys.}\ }\textbf {\bibinfo {volume} {70}},\ \bibinfo
  {pages} {1039} (\bibinfo {year} {1998})}\BibitemShut {NoStop}%
\bibitem [{\citenamefont {Tokura}(2006)}]{Tokura}%
  \BibitemOpen
  \bibfield  {author} {\bibinfo {author} {\bibfnamefont {Y.}~\bibnamefont
  {Tokura}},\ }\href {\doibase 10.1088/0034-4885/69/3/R06} {\bibfield
  {journal} {\bibinfo  {journal} {Rep. Prog. Phys.}\ }\textbf {\bibinfo
  {volume} {69}},\ \bibinfo {pages} {797} (\bibinfo {year} {2006})}\BibitemShut
  {NoStop}%
\bibitem [{\citenamefont {Tachibana}\ \emph {et~al.}(2008)\citenamefont
  {Tachibana}, \citenamefont {Yoshida}, \citenamefont {Kawaji}, \citenamefont
  {Atake},\ and\ \citenamefont {Takayama-Muromachi}}]{Tachibana}%
  \BibitemOpen
  \bibfield  {author} {\bibinfo {author} {\bibfnamefont {M.}~\bibnamefont
  {Tachibana}}, \bibinfo {author} {\bibfnamefont {T.}~\bibnamefont {Yoshida}},
  \bibinfo {author} {\bibfnamefont {H.}~\bibnamefont {Kawaji}}, \bibinfo
  {author} {\bibfnamefont {T.}~\bibnamefont {Atake}}, \ and\ \bibinfo {author}
  {\bibfnamefont {E.}~\bibnamefont {Takayama-Muromachi}},\ }\href {\doibase
  10.1103/PhysRevB.77.094402} {\bibfield  {journal} {\bibinfo  {journal} {Phys.
  Rev. B}\ }\textbf {\bibinfo {volume} {77}},\ \bibinfo {pages} {094402}
  (\bibinfo {year} {2008})}\BibitemShut {NoStop}%
\bibitem [{\citenamefont {Goodenough}(1958)}]{Goodenough1958}%
  \BibitemOpen
  \bibfield  {author} {\bibinfo {author} {\bibfnamefont {J.~B.}\ \bibnamefont
  {Goodenough}},\ }\href {\doibase 10.1016/0022-3697(58)90107-0} {\bibfield
  {journal} {\bibinfo  {journal} {J. Phys. and Chem. of Solids}\ }\textbf
  {\bibinfo {volume} {6}},\ \bibinfo {pages} {287} (\bibinfo {year}
  {1958})}\BibitemShut {NoStop}%
\bibitem [{\citenamefont {Korotin}\ \emph {et~al.}(1996)\citenamefont
  {Korotin}, \citenamefont {Ezhov}, \citenamefont {Solovyev}, \citenamefont
  {Anisimov}, \citenamefont {Khomskii},\ and\ \citenamefont
  {Sawatzky}}]{Korotin}%
  \BibitemOpen
  \bibfield  {author} {\bibinfo {author} {\bibfnamefont {M.~A.}\ \bibnamefont
  {Korotin}}, \bibinfo {author} {\bibfnamefont {S.~Y.}\ \bibnamefont {Ezhov}},
  \bibinfo {author} {\bibfnamefont {I.~V.}\ \bibnamefont {Solovyev}}, \bibinfo
  {author} {\bibfnamefont {V.~I.}\ \bibnamefont {Anisimov}}, \bibinfo {author}
  {\bibfnamefont {D.~I.}\ \bibnamefont {Khomskii}}, \ and\ \bibinfo {author}
  {\bibfnamefont {G.~A.}\ \bibnamefont {Sawatzky}},\ }\href {\doibase
  10.1103/PhysRevB.54.5309} {\bibfield  {journal} {\bibinfo  {journal} {Phys.
  Rev. B}\ }\textbf {\bibinfo {volume} {54}},\ \bibinfo {pages} {5309}
  (\bibinfo {year} {1996})}\BibitemShut {NoStop}%
\bibitem [{\citenamefont {Raccah}\ and\ \citenamefont
  {Goodenough}(1967)}]{Raccah}%
  \BibitemOpen
  \bibfield  {author} {\bibinfo {author} {\bibfnamefont {P.~M.}\ \bibnamefont
  {Raccah}}\ and\ \bibinfo {author} {\bibfnamefont {J.~B.}\ \bibnamefont
  {Goodenough}},\ }\href {\doibase 10.1103/PhysRev.155.932} {\bibfield
  {journal} {\bibinfo  {journal} {Phys. Rev.}\ }\textbf {\bibinfo {volume}
  {155}},\ \bibinfo {pages} {932} (\bibinfo {year} {1967})}\BibitemShut
  {NoStop}%
\bibitem [{\citenamefont {Haverkort}\ \emph {et~al.}(2006)\citenamefont
  {Haverkort}, \citenamefont {Hu}, \citenamefont {Cezar}, \citenamefont
  {Burnus}, \citenamefont {Hartmann}, \citenamefont {Reuther}, \citenamefont
  {Zobel}, \citenamefont {Lorenz}, \citenamefont {Tanaka}, \citenamefont
  {Brookes}, \citenamefont {Hsieh}, \citenamefont {Lin}, \citenamefont {Chen},\
  and\ \citenamefont {Tjeng}}]{Haverkort}%
  \BibitemOpen
  \bibfield  {author} {\bibinfo {author} {\bibfnamefont {M.~W.}\ \bibnamefont
  {Haverkort}}, \bibinfo {author} {\bibfnamefont {Z.}~\bibnamefont {Hu}},
  \bibinfo {author} {\bibfnamefont {J.~C.}\ \bibnamefont {Cezar}}, \bibinfo
  {author} {\bibfnamefont {T.}~\bibnamefont {Burnus}}, \bibinfo {author}
  {\bibfnamefont {H.}~\bibnamefont {Hartmann}}, \bibinfo {author}
  {\bibfnamefont {M.}~\bibnamefont {Reuther}}, \bibinfo {author} {\bibfnamefont
  {C.}~\bibnamefont {Zobel}}, \bibinfo {author} {\bibfnamefont
  {T.}~\bibnamefont {Lorenz}}, \bibinfo {author} {\bibfnamefont
  {A.}~\bibnamefont {Tanaka}}, \bibinfo {author} {\bibfnamefont {N.~B.}\
  \bibnamefont {Brookes}}, \bibinfo {author} {\bibfnamefont {H.~H.}\
  \bibnamefont {Hsieh}}, \bibinfo {author} {\bibfnamefont {H.~J.}\ \bibnamefont
  {Lin}}, \bibinfo {author} {\bibfnamefont {C.~T.}\ \bibnamefont {Chen}}, \
  and\ \bibinfo {author} {\bibfnamefont {L.~H.}\ \bibnamefont {Tjeng}},\ }\href
  {\doibase 10.1103/PhysRevLett.97.176405} {\bibfield  {journal} {\bibinfo
  {journal} {Phys. Rev. Lett.}\ }\textbf {\bibinfo {volume} {97}},\ \bibinfo
  {pages} {176405} (\bibinfo {year} {2006})}\BibitemShut {NoStop}%
\bibitem [{\citenamefont {Kn\'{i}\v{z}ek}\ \emph {et~al.}(2009)\citenamefont
  {Kn\'{i}\v{z}ek}, \citenamefont {Jir\'{a}k}, \citenamefont {Hejtm\'{a}nek},
  \citenamefont {Nov\'{a}k},\ and\ \citenamefont {Ku}}]{Knizek}%
  \BibitemOpen
  \bibfield  {author} {\bibinfo {author} {\bibfnamefont {K.}~\bibnamefont
  {Kn\'{i}\v{z}ek}}, \bibinfo {author} {\bibfnamefont {Z.}~\bibnamefont
  {Jir\'{a}k}}, \bibinfo {author} {\bibfnamefont {J.}~\bibnamefont
  {Hejtm\'{a}nek}}, \bibinfo {author} {\bibfnamefont {P.}~\bibnamefont
  {Nov\'{a}k}}, \ and\ \bibinfo {author} {\bibfnamefont {W.}~\bibnamefont
  {Ku}},\ }\href {\doibase 10.1103/PhysRevB.79.014430} {\bibfield  {journal}
  {\bibinfo  {journal} {Phys. Rev. B}\ }\textbf {\bibinfo {volume} {79}},\
  \bibinfo {pages} {014430} (\bibinfo {year} {2009})}\BibitemShut {NoStop}%
\bibitem [{\citenamefont {Ropka}\ and\ \citenamefont
  {Radwanski}(2003)}]{Ropka2003}%
  \BibitemOpen
  \bibfield  {author} {\bibinfo {author} {\bibfnamefont {Z.}~\bibnamefont
  {Ropka}}\ and\ \bibinfo {author} {\bibfnamefont {R.~J.}\ \bibnamefont
  {Radwanski}},\ }\href {\doibase 10.1103/PhysRevB.67.172401} {\bibfield
  {journal} {\bibinfo  {journal} {Phys. Rev. B}\ }\textbf {\bibinfo {volume}
  {67}},\ \bibinfo {pages} {172401} (\bibinfo {year} {2003})}\BibitemShut
  {NoStop}%
\bibitem [{\citenamefont {Ky\^{o}men}\ \emph {et~al.}(2003)\citenamefont
  {Ky\^{o}men}, \citenamefont {Asaka},\ and\ \citenamefont
  {Itoh}}]{Kyomen2003}%
  \BibitemOpen
  \bibfield  {author} {\bibinfo {author} {\bibfnamefont {T.}~\bibnamefont
  {Ky\^{o}men}}, \bibinfo {author} {\bibfnamefont {Y.}~\bibnamefont {Asaka}}, \
  and\ \bibinfo {author} {\bibfnamefont {M.}~\bibnamefont {Itoh}},\ }\href
  {\doibase 10.1103/PhysRevB.67.144424} {\bibfield  {journal} {\bibinfo
  {journal} {Phys. Rev. B}\ }\textbf {\bibinfo {volume} {67}},\ \bibinfo
  {pages} {144424} (\bibinfo {year} {2003})}\BibitemShut {NoStop}%
\bibitem [{\citenamefont {Ky\^{o}men}\ \emph {et~al.}(2005)\citenamefont
  {Ky\^{o}men}, \citenamefont {Asaka},\ and\ \citenamefont
  {Itoh}}]{Kyomen2005}%
  \BibitemOpen
  \bibfield  {author} {\bibinfo {author} {\bibfnamefont {T.}~\bibnamefont
  {Ky\^{o}men}}, \bibinfo {author} {\bibfnamefont {Y.}~\bibnamefont {Asaka}}, \
  and\ \bibinfo {author} {\bibfnamefont {M.}~\bibnamefont {Itoh}},\ }\href
  {\doibase 10.1103/PhysRevB.71.024418} {\bibfield  {journal} {\bibinfo
  {journal} {Phys. Rev. B}\ }\textbf {\bibinfo {volume} {71}},\ \bibinfo
  {pages} {024418} (\bibinfo {year} {2005})}\BibitemShut {NoStop}%
\bibitem [{\citenamefont {Yamaguchi}\ \emph {et~al.}(1997)\citenamefont
  {Yamaguchi}, \citenamefont {Okimoto},\ and\ \citenamefont
  {Tokura}}]{Yamaguchi1997}%
  \BibitemOpen
  \bibfield  {author} {\bibinfo {author} {\bibfnamefont {S.}~\bibnamefont
  {Yamaguchi}}, \bibinfo {author} {\bibfnamefont {Y.}~\bibnamefont {Okimoto}},
  \ and\ \bibinfo {author} {\bibfnamefont {Y.}~\bibnamefont {Tokura}},\ }\href
  {\doibase 10.1103/PhysRevB.55.R8666} {\bibfield  {journal} {\bibinfo
  {journal} {Phys. Rev. B}\ }\textbf {\bibinfo {volume} {55}},\ \bibinfo
  {pages} {R8666} (\bibinfo {year} {1997})}\BibitemShut {NoStop}%
\bibitem [{\citenamefont {Ishikawa}\ \emph {et~al.}(2004)\citenamefont
  {Ishikawa}, \citenamefont {Nohara},\ and\ \citenamefont {Sugai}}]{Ishikawa}%
  \BibitemOpen
  \bibfield  {author} {\bibinfo {author} {\bibfnamefont {A.}~\bibnamefont
  {Ishikawa}}, \bibinfo {author} {\bibfnamefont {J.}~\bibnamefont {Nohara}}, \
  and\ \bibinfo {author} {\bibfnamefont {S.}~\bibnamefont {Sugai}},\ }\href
  {\doibase 10.1103/PhysRevLett.93.136401} {\bibfield  {journal} {\bibinfo
  {journal} {Phys. Rev. Lett.}\ }\textbf {\bibinfo {volume} {93}},\ \bibinfo
  {pages} {136401} (\bibinfo {year} {2004})}\BibitemShut {NoStop}%
\bibitem [{\citenamefont {Kune\v{s}}\ and\ \citenamefont
  {K\v{r}\'{a}pek}(2011)}]{Kunes2011}%
  \BibitemOpen
  \bibfield  {author} {\bibinfo {author} {\bibfnamefont {J.}~\bibnamefont
  {Kune\v{s}}}\ and\ \bibinfo {author} {\bibfnamefont {V.}~\bibnamefont
  {K\v{r}\'{a}pek}},\ }\href {\doibase 10.1103/PhysRevLett.106.256401}
  {\bibfield  {journal} {\bibinfo  {journal} {Phys. Rev. Lett.}\ }\textbf
  {\bibinfo {volume} {106}},\ \bibinfo {pages} {256401} (\bibinfo {year}
  {2011})}\BibitemShut {NoStop}%
\bibitem [{\citenamefont {Kanamori}\ \emph {et~al.}(2011)\citenamefont
  {Kanamori}, \citenamefont {Matsueda},\ and\ \citenamefont
  {Ishihara}}]{Kanamori2011}%
  \BibitemOpen
  \bibfield  {author} {\bibinfo {author} {\bibfnamefont {Y.}~\bibnamefont
  {Kanamori}}, \bibinfo {author} {\bibfnamefont {H.}~\bibnamefont {Matsueda}},
  \ and\ \bibinfo {author} {\bibfnamefont {S.}~\bibnamefont {Ishihara}},\
  }\href {\doibase 10.1103/PhysRevLett.107.167403} {\bibfield  {journal}
  {\bibinfo  {journal} {Phys. Rev. Lett.}\ }\textbf {\bibinfo {volume} {107}},\
  \bibinfo {pages} {167403} (\bibinfo {year} {2011})}\BibitemShut {NoStop}%
\bibitem [{\citenamefont {K\v{r}\'{a}pek}\ \emph {et~al.}(2012)\citenamefont
  {K\v{r}\'{a}pek}, \citenamefont {Nov\'{a}k}, \citenamefont {Kune\v{s}},
  \citenamefont {Novoselov}, \citenamefont {Korotin},\ and\ \citenamefont
  {Anisimov}}]{Krapek}%
  \BibitemOpen
  \bibfield  {author} {\bibinfo {author} {\bibfnamefont {V.}~\bibnamefont
  {K\v{r}\'{a}pek}}, \bibinfo {author} {\bibfnamefont {P.}~\bibnamefont
  {Nov\'{a}k}}, \bibinfo {author} {\bibfnamefont {J.}~\bibnamefont
  {Kune\v{s}}}, \bibinfo {author} {\bibfnamefont {D.}~\bibnamefont
  {Novoselov}}, \bibinfo {author} {\bibfnamefont {D.~M.}\ \bibnamefont
  {Korotin}}, \ and\ \bibinfo {author} {\bibfnamefont {V.~I.}\ \bibnamefont
  {Anisimov}},\ }\href {\doibase 10.1103/PhysRevB.86.195104} {\bibfield
  {journal} {\bibinfo  {journal} {Phys. Rev. B}\ }\textbf {\bibinfo {volume}
  {86}},\ \bibinfo {pages} {195104} (\bibinfo {year} {2012})}\BibitemShut
  {NoStop}%
\bibitem [{\citenamefont {Doi}\ \emph {et~al.}(2014)\citenamefont {Doi},
  \citenamefont {Fujioka}, \citenamefont {Fukuda}, \citenamefont {Tsutsui},
  \citenamefont {Okuyama}, \citenamefont {Taguchi}, \citenamefont {Arima},
  \citenamefont {Baron},\ and\ \citenamefont {Tokura}}]{Doi}%
  \BibitemOpen
  \bibfield  {author} {\bibinfo {author} {\bibfnamefont {A.}~\bibnamefont
  {Doi}}, \bibinfo {author} {\bibfnamefont {J.}~\bibnamefont {Fujioka}},
  \bibinfo {author} {\bibfnamefont {T.}~\bibnamefont {Fukuda}}, \bibinfo
  {author} {\bibfnamefont {S.}~\bibnamefont {Tsutsui}}, \bibinfo {author}
  {\bibfnamefont {D.}~\bibnamefont {Okuyama}}, \bibinfo {author} {\bibfnamefont
  {Y.}~\bibnamefont {Taguchi}}, \bibinfo {author} {\bibfnamefont
  {T.}~\bibnamefont {Arima}}, \bibinfo {author} {\bibfnamefont {A.~Q.~R.}\
  \bibnamefont {Baron}}, \ and\ \bibinfo {author} {\bibfnamefont
  {Y.}~\bibnamefont {Tokura}},\ }\href {\doibase 10.1103/PhysRevB.90.081109}
  {\bibfield  {journal} {\bibinfo  {journal} {Phys. Rev. B}\ }\textbf {\bibinfo
  {volume} {90}},\ \bibinfo {pages} {081109} (\bibinfo {year}
  {2014})}\BibitemShut {NoStop}%
\bibitem [{\citenamefont {Tokunaga}\ \emph {et~al.}(1998)\citenamefont
  {Tokunaga}, \citenamefont {Miura}, \citenamefont {Tomioka},\ and\
  \citenamefont {Tokura}}]{Tokunaga}%
  \BibitemOpen
  \bibfield  {author} {\bibinfo {author} {\bibfnamefont {M.}~\bibnamefont
  {Tokunaga}}, \bibinfo {author} {\bibfnamefont {N.}~\bibnamefont {Miura}},
  \bibinfo {author} {\bibfnamefont {Y.}~\bibnamefont {Tomioka}}, \ and\
  \bibinfo {author} {\bibfnamefont {Y.}~\bibnamefont {Tokura}},\ }\href
  {\doibase 10.1103/PhysRevB.57.5259} {\bibfield  {journal} {\bibinfo
  {journal} {Phys. Rev. B}\ }\textbf {\bibinfo {volume} {57}},\ \bibinfo
  {pages} {5259} (\bibinfo {year} {1998})}\BibitemShut {NoStop}%
\bibitem [{\citenamefont {Nomura}\ \emph {et~al.}(2015)\citenamefont {Nomura},
  \citenamefont {Matsuda}, \citenamefont {Takeyama}, \citenamefont {Matsuo},
  \citenamefont {Kindo},\ and\ \citenamefont {Kobayashi}}]{Nomura}%
  \BibitemOpen
  \bibfield  {author} {\bibinfo {author} {\bibfnamefont {T.}~\bibnamefont
  {Nomura}}, \bibinfo {author} {\bibfnamefont {Y.~H.}\ \bibnamefont {Matsuda}},
  \bibinfo {author} {\bibfnamefont {S.}~\bibnamefont {Takeyama}}, \bibinfo
  {author} {\bibfnamefont {A.}~\bibnamefont {Matsuo}}, \bibinfo {author}
  {\bibfnamefont {K.}~\bibnamefont {Kindo}}, \ and\ \bibinfo {author}
  {\bibfnamefont {T.~C.}\ \bibnamefont {Kobayashi}},\ }\href {\doibase
  10.1103/PhysRevB.92.064109} {\bibfield  {journal} {\bibinfo  {journal} {Phys.
  Rev. B}\ }\textbf {\bibinfo {volume} {92}},\ \bibinfo {pages} {064109}
  (\bibinfo {year} {2015})}\BibitemShut {NoStop}%
\bibitem [{\citenamefont {Yamaguchi}\ \emph {et~al.}(1996)\citenamefont
  {Yamaguchi}, \citenamefont {Okimoto}, \citenamefont {Taniguchi},\ and\
  \citenamefont {Tokura}}]{Yamaguchi1996}%
  \BibitemOpen
  \bibfield  {author} {\bibinfo {author} {\bibfnamefont {S.}~\bibnamefont
  {Yamaguchi}}, \bibinfo {author} {\bibfnamefont {Y.}~\bibnamefont {Okimoto}},
  \bibinfo {author} {\bibfnamefont {H.}~\bibnamefont {Taniguchi}}, \ and\
  \bibinfo {author} {\bibfnamefont {Y.}~\bibnamefont {Tokura}},\ }\href
  {\doibase 10.1103/PhysRevB.53.R2926} {\bibfield  {journal} {\bibinfo
  {journal} {Phys. Rev. B}\ }\textbf {\bibinfo {volume} {53}},\ \bibinfo
  {pages} {R2926} (\bibinfo {year} {1996})}\BibitemShut {NoStop}%
\bibitem [{\citenamefont {Sato}\ \emph {et~al.}(2009)\citenamefont {Sato},
  \citenamefont {Matsuo}, \citenamefont {Kindo}, \citenamefont {Kobayashi},\
  and\ \citenamefont {Asai}}]{Sato2009}%
  \BibitemOpen
  \bibfield  {author} {\bibinfo {author} {\bibfnamefont {K.}~\bibnamefont
  {Sato}}, \bibinfo {author} {\bibfnamefont {A.}~\bibnamefont {Matsuo}},
  \bibinfo {author} {\bibfnamefont {K.}~\bibnamefont {Kindo}}, \bibinfo
  {author} {\bibfnamefont {Y.}~\bibnamefont {Kobayashi}}, \ and\ \bibinfo
  {author} {\bibfnamefont {K.}~\bibnamefont {Asai}},\ }\href {\doibase
  10.1143/JPSJ.80.104702} {\bibfield  {journal} {\bibinfo  {journal} {J. Phys.
  Soc. Jpn.}\ }\textbf {\bibinfo {volume} {78}},\ \bibinfo {pages} {093702}
  (\bibinfo {year} {2009})}\BibitemShut {NoStop}%
\bibitem [{\citenamefont {Altarawneh}\ \emph {et~al.}(2012)\citenamefont
  {Altarawneh}, \citenamefont {Chern}, \citenamefont {Harrison}, \citenamefont
  {Batista}, \citenamefont {Uchida}, \citenamefont {Jaime}, \citenamefont
  {Rickel}, \citenamefont {Crooker}, \citenamefont {Mielke}, \citenamefont
  {Betts}, \citenamefont {Mitchell},\ and\ \citenamefont {Hoch}}]{Moaz}%
  \BibitemOpen
  \bibfield  {author} {\bibinfo {author} {\bibfnamefont {M.~M.}\ \bibnamefont
  {Altarawneh}}, \bibinfo {author} {\bibfnamefont {G.~W.}\ \bibnamefont
  {Chern}}, \bibinfo {author} {\bibfnamefont {N.}~\bibnamefont {Harrison}},
  \bibinfo {author} {\bibfnamefont {C.~D.}\ \bibnamefont {Batista}}, \bibinfo
  {author} {\bibfnamefont {A.}~\bibnamefont {Uchida}}, \bibinfo {author}
  {\bibfnamefont {M.}~\bibnamefont {Jaime}}, \bibinfo {author} {\bibfnamefont
  {D.~G.}\ \bibnamefont {Rickel}}, \bibinfo {author} {\bibfnamefont {S.~A.}\
  \bibnamefont {Crooker}}, \bibinfo {author} {\bibfnamefont {C.~H.}\
  \bibnamefont {Mielke}}, \bibinfo {author} {\bibfnamefont {J.~B.}\
  \bibnamefont {Betts}}, \bibinfo {author} {\bibfnamefont {J.~F.}\ \bibnamefont
  {Mitchell}}, \ and\ \bibinfo {author} {\bibfnamefont {M.~J.~R.}\ \bibnamefont
  {Hoch}},\ }\href {\doibase 10.1103/PhysRevLett.109.037201} {\bibfield
  {journal} {\bibinfo  {journal} {Phys. Rev. Lett.}\ }\textbf {\bibinfo
  {volume} {109}},\ \bibinfo {pages} {037201} (\bibinfo {year}
  {2012})}\BibitemShut {NoStop}%
\bibitem [{\citenamefont {Rotter}\ \emph {et~al.}(2014)\citenamefont {Rotter},
  \citenamefont {Wang}, \citenamefont {Boothroyd}, \citenamefont {Prabhakaran},
  \citenamefont {Tanaka},\ and\ \citenamefont {Doerr}}]{Rotter}%
  \BibitemOpen
  \bibfield  {author} {\bibinfo {author} {\bibfnamefont {M.}~\bibnamefont
  {Rotter}}, \bibinfo {author} {\bibfnamefont {Z.~S.}\ \bibnamefont {Wang}},
  \bibinfo {author} {\bibfnamefont {A.~T.}\ \bibnamefont {Boothroyd}}, \bibinfo
  {author} {\bibfnamefont {D.}~\bibnamefont {Prabhakaran}}, \bibinfo {author}
  {\bibfnamefont {A.}~\bibnamefont {Tanaka}}, \ and\ \bibinfo {author}
  {\bibfnamefont {M.}~\bibnamefont {Doerr}},\ }\href {\doibase
  10.1038/srep07003} {\bibfield  {journal} {\bibinfo  {journal} {Sci. Rep.}\
  }\textbf {\bibinfo {volume} {4}},\ \bibinfo {pages} {7003} (\bibinfo {year}
  {2014})}\BibitemShut {NoStop}%
\bibitem [{\citenamefont {Platonov}\ \emph {et~al.}(2012)\citenamefont
  {Platonov}, \citenamefont {Kudasov}, \citenamefont {Monakhov},\ and\
  \citenamefont {Tatsenko}}]{Platonov}%
  \BibitemOpen
  \bibfield  {author} {\bibinfo {author} {\bibfnamefont {V.~V.}\ \bibnamefont
  {Platonov}}, \bibinfo {author} {\bibfnamefont {Y.~B.}\ \bibnamefont
  {Kudasov}}, \bibinfo {author} {\bibfnamefont {M.~P.}\ \bibnamefont
  {Monakhov}}, \ and\ \bibinfo {author} {\bibfnamefont {O.~M.}\ \bibnamefont
  {Tatsenko}},\ }\href {\doibase 10.1134/S1063783412020266} {\bibfield
  {journal} {\bibinfo  {journal} {Phys. Solid State}\ }\textbf {\bibinfo
  {volume} {54}},\ \bibinfo {pages} {279} (\bibinfo {year} {2012})}\BibitemShut
  {NoStop}%
\bibitem [{\citenamefont {Hoch}\ \emph {et~al.}(2009)\citenamefont {Hoch},
  \citenamefont {Nellutla}, \citenamefont {van Tol}, \citenamefont {Choi},
  \citenamefont {Lu}, \citenamefont {Zheng},\ and\ \citenamefont
  {Mitchell}}]{Hoch}%
  \BibitemOpen
  \bibfield  {author} {\bibinfo {author} {\bibfnamefont {M.~J.~R.}\
  \bibnamefont {Hoch}}, \bibinfo {author} {\bibfnamefont {S.}~\bibnamefont
  {Nellutla}}, \bibinfo {author} {\bibfnamefont {J.}~\bibnamefont {van Tol}},
  \bibinfo {author} {\bibfnamefont {E.~S.}\ \bibnamefont {Choi}}, \bibinfo
  {author} {\bibfnamefont {J.}~\bibnamefont {Lu}}, \bibinfo {author}
  {\bibfnamefont {H.}~\bibnamefont {Zheng}}, \ and\ \bibinfo {author}
  {\bibfnamefont {J.~F.}\ \bibnamefont {Mitchell}},\ }\href {\doibase
  10.1103/PhysRevB.79.214421} {\bibfield  {journal} {\bibinfo  {journal} {Phys.
  Rev. B}\ }\textbf {\bibinfo {volume} {79}},\ \bibinfo {pages} {214421}
  (\bibinfo {year} {2009})}\BibitemShut {NoStop}%
\bibitem [{\citenamefont {Miura}\ \emph {et~al.}(2003)\citenamefont {Miura},
  \citenamefont {Osada},\ and\ \citenamefont {Takeyama}}]{Miura}%
  \BibitemOpen
  \bibfield  {author} {\bibinfo {author} {\bibfnamefont {N.}~\bibnamefont
  {Miura}}, \bibinfo {author} {\bibfnamefont {T.}~\bibnamefont {Osada}}, \ and\
  \bibinfo {author} {\bibfnamefont {S.}~\bibnamefont {Takeyama}},\ }\href
  {\doibase 10.1023/A:1025689218138} {\bibfield  {journal} {\bibinfo  {journal}
  {J. Low Temp. Phys.}\ }\textbf {\bibinfo {volume} {133}},\ \bibinfo {pages}
  {139} (\bibinfo {year} {2003})}\BibitemShut {NoStop}%
\bibitem [{\citenamefont {Takeyama}\ \emph {et~al.}(1988)\citenamefont
  {Takeyama}, \citenamefont {Amaya}, \citenamefont {Nakagawa}, \citenamefont
  {Ishizuka}, \citenamefont {Nakao}, \citenamefont {Sakakibara}, \citenamefont
  {Gotot}, \citenamefont {Miura}, \citenamefont {Ajiro},\ and\ \citenamefont
  {Kikuchi}}]{Takeyama1988}%
  \BibitemOpen
  \bibfield  {author} {\bibinfo {author} {\bibfnamefont {S.}~\bibnamefont
  {Takeyama}}, \bibinfo {author} {\bibfnamefont {K.}~\bibnamefont {Amaya}},
  \bibinfo {author} {\bibfnamefont {T.}~\bibnamefont {Nakagawa}}, \bibinfo
  {author} {\bibfnamefont {M.}~\bibnamefont {Ishizuka}}, \bibinfo {author}
  {\bibfnamefont {K.}~\bibnamefont {Nakao}}, \bibinfo {author} {\bibfnamefont
  {T.}~\bibnamefont {Sakakibara}}, \bibinfo {author} {\bibfnamefont
  {T.}~\bibnamefont {Gotot}}, \bibinfo {author} {\bibfnamefont
  {N.}~\bibnamefont {Miura}}, \bibinfo {author} {\bibfnamefont
  {Y.}~\bibnamefont {Ajiro}}, \ and\ \bibinfo {author} {\bibfnamefont
  {H.}~\bibnamefont {Kikuchi}},\ }\href {\doibase 10.1088/0022-3735/21/11/004}
  {\bibfield  {journal} {\bibinfo  {journal} {J. Phys. E}\ }\textbf {\bibinfo
  {volume} {21}},\ \bibinfo {pages} {1025} (\bibinfo {year}
  {1988})}\BibitemShut {NoStop}%
\bibitem [{\citenamefont {Amaya}\ \emph {et~al.}(1989)\citenamefont {Amaya},
  \citenamefont {Takeyama}, \citenamefont {Nakagawa}, \citenamefont {Ishizuka},
  \citenamefont {Nakao}, \citenamefont {Sakakibara}, \citenamefont {Goto},
  \citenamefont {Miura}, \citenamefont {Ajiro},\ and\ \citenamefont
  {Kikuchi}}]{Amaya}%
  \BibitemOpen
  \bibfield  {author} {\bibinfo {author} {\bibfnamefont {K.}~\bibnamefont
  {Amaya}}, \bibinfo {author} {\bibfnamefont {S.}~\bibnamefont {Takeyama}},
  \bibinfo {author} {\bibfnamefont {T.}~\bibnamefont {Nakagawa}}, \bibinfo
  {author} {\bibfnamefont {M.}~\bibnamefont {Ishizuka}}, \bibinfo {author}
  {\bibfnamefont {K.}~\bibnamefont {Nakao}}, \bibinfo {author} {\bibfnamefont
  {T.}~\bibnamefont {Sakakibara}}, \bibinfo {author} {\bibfnamefont
  {T.}~\bibnamefont {Goto}}, \bibinfo {author} {\bibfnamefont {N.}~\bibnamefont
  {Miura}}, \bibinfo {author} {\bibfnamefont {Y.}~\bibnamefont {Ajiro}}, \ and\
  \bibinfo {author} {\bibfnamefont {H.}~\bibnamefont {Kikuchi}},\ }\href
  {\doibase 10.1016/0921-4526(89)90538-3} {\bibfield  {journal} {\bibinfo
  {journal} {Physica B}\ }\textbf {\bibinfo {volume} {155}},\ \bibinfo {pages}
  {396} (\bibinfo {year} {1989})}\BibitemShut {NoStop}%
\bibitem [{\citenamefont {Takeyama}\ \emph {et~al.}(2012)\citenamefont
  {Takeyama}, \citenamefont {Sakakura}, \citenamefont {Matsuda}, \citenamefont
  {Miyata},\ and\ \citenamefont {Tokunaga}}]{Takeyama2012}%
  \BibitemOpen
  \bibfield  {author} {\bibinfo {author} {\bibfnamefont {S.}~\bibnamefont
  {Takeyama}}, \bibinfo {author} {\bibfnamefont {R.}~\bibnamefont {Sakakura}},
  \bibinfo {author} {\bibfnamefont {Y.~H.}\ \bibnamefont {Matsuda}}, \bibinfo
  {author} {\bibfnamefont {A.}~\bibnamefont {Miyata}}, \ and\ \bibinfo {author}
  {\bibfnamefont {M.}~\bibnamefont {Tokunaga}},\ }\href {\doibase
  10.1143/JPSJ.81.014702} {\bibfield  {journal} {\bibinfo  {journal} {J. Phys.
  Soc. Jpn.}\ }\textbf {\bibinfo {volume} {81}},\ \bibinfo {pages} {014702}
  (\bibinfo {year} {2012})}\BibitemShut {NoStop}%
\bibitem [{\citenamefont {Landau}\ and\ \citenamefont
  {Lifshitz}(1980)}]{Landau}%
  \BibitemOpen
  \bibfield  {author} {\bibinfo {author} {\bibfnamefont {L.~D.}\ \bibnamefont
  {Landau}}\ and\ \bibinfo {author} {\bibfnamefont {E.~M.}\ \bibnamefont
  {Lifshitz}},\ }\href@noop {} {\emph {\bibinfo {title} {\textit{Statistical
  Physics} 3rd Ed. Part 1}}}\ (\bibinfo  {publisher} {Elsevier},\ \bibinfo
  {address} {Amsterdam},\ \bibinfo {year} {1980})\BibitemShut {NoStop}%
\bibitem [{\citenamefont {St{\o}len}\ \emph {et~al.}(1997)\citenamefont
  {St{\o}len}, \citenamefont {Gr{\o}nvold}, \citenamefont {Brinks},
  \citenamefont {Atake},\ and\ \citenamefont {Mori}}]{stolen}%
  \BibitemOpen
  \bibfield  {author} {\bibinfo {author} {\bibfnamefont {S.}~\bibnamefont
  {St{\o}len}}, \bibinfo {author} {\bibfnamefont {F.}~\bibnamefont
  {Gr{\o}nvold}}, \bibinfo {author} {\bibfnamefont {H.}~\bibnamefont {Brinks}},
  \bibinfo {author} {\bibfnamefont {T.}~\bibnamefont {Atake}}, \ and\ \bibinfo
  {author} {\bibfnamefont {H.}~\bibnamefont {Mori}},\ }\href {\doibase
  10.1103/PhysRevB.55.14103} {\bibfield  {journal} {\bibinfo  {journal} {Phys.
  Rev. B}\ }\textbf {\bibinfo {volume} {55}},\ \bibinfo {pages} {14103}
  (\bibinfo {year} {1997})}\BibitemShut {NoStop}%
\bibitem [{\citenamefont {Biernacki}\ and\ \citenamefont
  {Clerjaud}(2005)}]{Biernacki2005}%
  \BibitemOpen
  \bibfield  {author} {\bibinfo {author} {\bibfnamefont {S.~W.}\ \bibnamefont
  {Biernacki}}\ and\ \bibinfo {author} {\bibfnamefont {B.}~\bibnamefont
  {Clerjaud}},\ }\href {\doibase 10.1103/PhysRevB.72.024406} {\bibfield
  {journal} {\bibinfo  {journal} {Phys. Rev. B}\ }\textbf {\bibinfo {volume}
  {72}},\ \bibinfo {pages} {024406} (\bibinfo {year} {2005})}\BibitemShut
  {NoStop}%
\bibitem [{\citenamefont {Kimura}\ \emph {et~al.}(2008)\citenamefont {Kimura},
  \citenamefont {Maeda}, \citenamefont {Kashiwagi}, \citenamefont {Yamaguchi},
  \citenamefont {Hagiwara}, \citenamefont {Yoshida}, \citenamefont {Terasaki},\
  and\ \citenamefont {Kindo}}]{Kimura2008}%
  \BibitemOpen
  \bibfield  {author} {\bibinfo {author} {\bibfnamefont {S.}~\bibnamefont
  {Kimura}}, \bibinfo {author} {\bibfnamefont {Y.}~\bibnamefont {Maeda}},
  \bibinfo {author} {\bibfnamefont {T.}~\bibnamefont {Kashiwagi}}, \bibinfo
  {author} {\bibfnamefont {H.}~\bibnamefont {Yamaguchi}}, \bibinfo {author}
  {\bibfnamefont {M.}~\bibnamefont {Hagiwara}}, \bibinfo {author}
  {\bibfnamefont {S.}~\bibnamefont {Yoshida}}, \bibinfo {author} {\bibfnamefont
  {I.}~\bibnamefont {Terasaki}}, \ and\ \bibinfo {author} {\bibfnamefont
  {K.}~\bibnamefont {Kindo}},\ }\href {\doibase 10.1103/PhysRevB.78.180403}
  {\bibfield  {journal} {\bibinfo  {journal} {Phys. Rev. B}\ }\textbf {\bibinfo
  {volume} {78}},\ \bibinfo {pages} {180403} (\bibinfo {year}
  {2008})}\BibitemShut {NoStop}%
\bibitem [{\citenamefont {Mary\v{s}ko}\ \emph {et~al.}(2011)\citenamefont
  {Mary\v{s}ko}, \citenamefont {Jir\'{a}k}, \citenamefont {Kn\'{i}\v{z}ek},
  \citenamefont {Nov\'{a}k}, \citenamefont {Hejtm\'{a}nek}, \citenamefont
  {Naito}, \citenamefont {Sasaki},\ and\ \citenamefont {Fujishiro}}]{Marysko}%
  \BibitemOpen
  \bibfield  {author} {\bibinfo {author} {\bibfnamefont {M.}~\bibnamefont
  {Mary\v{s}ko}}, \bibinfo {author} {\bibfnamefont {Z.}~\bibnamefont
  {Jir\'{a}k}}, \bibinfo {author} {\bibfnamefont {K.}~\bibnamefont
  {Kn\'{i}\v{z}ek}}, \bibinfo {author} {\bibfnamefont {P.}~\bibnamefont
  {Nov\'{a}k}}, \bibinfo {author} {\bibfnamefont {J.}~\bibnamefont
  {Hejtm\'{a}nek}}, \bibinfo {author} {\bibfnamefont {T.}~\bibnamefont
  {Naito}}, \bibinfo {author} {\bibfnamefont {H.}~\bibnamefont {Sasaki}}, \
  and\ \bibinfo {author} {\bibfnamefont {H.}~\bibnamefont {Fujishiro}},\ }\href
  {\doibase 10.1063/1.3559485} {\bibfield  {journal} {\bibinfo  {journal} {J.
  Appl. Phys.}\ }\textbf {\bibinfo {volume} {109}},\ \bibinfo {pages} {07E127}
  (\bibinfo {year} {2011})}\BibitemShut {NoStop}%
\bibitem [{\citenamefont {Naito}\ \emph {et~al.}(2014)\citenamefont {Naito},
  \citenamefont {Fujishiro}, \citenamefont {Nishizaki}, \citenamefont
  {Kobayashi}, \citenamefont {Hejtm\'{a}nek}, \citenamefont {Kn\'{i}\v{z}ek},\
  and\ \citenamefont {Jir\'{a}k}}]{Naito2014}%
  \BibitemOpen
  \bibfield  {author} {\bibinfo {author} {\bibfnamefont {T.}~\bibnamefont
  {Naito}}, \bibinfo {author} {\bibfnamefont {H.}~\bibnamefont {Fujishiro}},
  \bibinfo {author} {\bibfnamefont {T.}~\bibnamefont {Nishizaki}}, \bibinfo
  {author} {\bibfnamefont {N.}~\bibnamefont {Kobayashi}}, \bibinfo {author}
  {\bibfnamefont {J.}~\bibnamefont {Hejtm\'{a}nek}}, \bibinfo {author}
  {\bibfnamefont {K.}~\bibnamefont {Kn\'{i}\v{z}ek}}, \ and\ \bibinfo {author}
  {\bibfnamefont {Z.}~\bibnamefont {Jir\'{a}k}},\ }\href {\doibase
  10.1063/1.4884435} {\bibfield  {journal} {\bibinfo  {journal} {J. Appl.
  Phys.}\ }\textbf {\bibinfo {volume} {115}},\ \bibinfo {pages} {233914}
  (\bibinfo {year} {2014})}\BibitemShut {NoStop}%
\bibitem [{\citenamefont {Qi}\ \emph {et~al.}(1983)\citenamefont {Qi},
  \citenamefont {M\"{u}ller}, \citenamefont {Spiering},\ and\ \citenamefont
  {G\"{u}tlich}}]{Qi}%
  \BibitemOpen
  \bibfield  {author} {\bibinfo {author} {\bibfnamefont {Y.}~\bibnamefont
  {Qi}}, \bibinfo {author} {\bibfnamefont {E.~W.}\ \bibnamefont {M\"{u}ller}},
  \bibinfo {author} {\bibfnamefont {H.}~\bibnamefont {Spiering}}, \ and\
  \bibinfo {author} {\bibfnamefont {P.}~\bibnamefont {G\"{u}tlich}},\ }\href
  {\doibase 10.1016/0009-2614(83)87521-6} {\bibfield  {journal} {\bibinfo
  {journal} {Chem. Phys. Lett.}\ }\textbf {\bibinfo {volume} {101}},\ \bibinfo
  {pages} {503} (\bibinfo {year} {1983})}\BibitemShut {NoStop}%
\bibitem [{\citenamefont {Kimura}\ \emph {et~al.}(2005)\citenamefont {Kimura},
  \citenamefont {Narumi}, \citenamefont {Kindo}, \citenamefont {Nakano},\ and\
  \citenamefont {Matsubayashi}}]{Kimura2005}%
  \BibitemOpen
  \bibfield  {author} {\bibinfo {author} {\bibfnamefont {S.}~\bibnamefont
  {Kimura}}, \bibinfo {author} {\bibfnamefont {Y.}~\bibnamefont {Narumi}},
  \bibinfo {author} {\bibfnamefont {K.}~\bibnamefont {Kindo}}, \bibinfo
  {author} {\bibfnamefont {M.}~\bibnamefont {Nakano}}, \ and\ \bibinfo {author}
  {\bibfnamefont {G.~E.}\ \bibnamefont {Matsubayashi}},\ }\href {\doibase
  10.1103/PhysRevB.72.064448} {\bibfield  {journal} {\bibinfo  {journal} {Phys.
  Rev. B}\ }\textbf {\bibinfo {volume} {72}},\ \bibinfo {pages} {064448}
  (\bibinfo {year} {2005})}\BibitemShut {NoStop}%
\bibitem [{\citenamefont {Asai}\ \emph {et~al.}(1994)\citenamefont {Asai},
  \citenamefont {Yokokura}, \citenamefont {Nishimori}, \citenamefont {Chou},
  \citenamefont {Tranquada}, \citenamefont {Shirane}, \citenamefont {Higuchi},
  \citenamefont {Okajima},\ and\ \citenamefont {Kohn}}]{Asai1994}%
  \BibitemOpen
  \bibfield  {author} {\bibinfo {author} {\bibfnamefont {K.}~\bibnamefont
  {Asai}}, \bibinfo {author} {\bibfnamefont {O.}~\bibnamefont {Yokokura}},
  \bibinfo {author} {\bibfnamefont {N.}~\bibnamefont {Nishimori}}, \bibinfo
  {author} {\bibfnamefont {H.}~\bibnamefont {Chou}}, \bibinfo {author}
  {\bibfnamefont {J.~M.}\ \bibnamefont {Tranquada}}, \bibinfo {author}
  {\bibfnamefont {G.}~\bibnamefont {Shirane}}, \bibinfo {author} {\bibfnamefont
  {S.}~\bibnamefont {Higuchi}}, \bibinfo {author} {\bibfnamefont
  {Y.}~\bibnamefont {Okajima}}, \ and\ \bibinfo {author} {\bibfnamefont
  {K.}~\bibnamefont {Kohn}},\ }\href {\doibase 10.1103/PhysRevB.50.3025}
  {\bibfield  {journal} {\bibinfo  {journal} {Phys. Rev. B}\ }\textbf {\bibinfo
  {volume} {50}},\ \bibinfo {pages} {3025} (\bibinfo {year}
  {1994})}\BibitemShut {NoStop}%
\bibitem [{\citenamefont {Kune\v{s}}\ and\ \citenamefont
  {Augustinsk\'{y}}(2014{\natexlab{a}})}]{Kunes2014}%
  \BibitemOpen
  \bibfield  {author} {\bibinfo {author} {\bibfnamefont {J.}~\bibnamefont
  {Kune\v{s}}}\ and\ \bibinfo {author} {\bibfnamefont {P.}~\bibnamefont
  {Augustinsk\'{y}}},\ }\href {\doibase DOI:10.1103/PhysRevB.89.115134}
  {\bibfield  {journal} {\bibinfo  {journal} {Phys. Rev. B}\ }\textbf {\bibinfo
  {volume} {89}},\ \bibinfo {pages} {115134} (\bibinfo {year}
  {2014}{\natexlab{a}})}\BibitemShut {NoStop}%
\bibitem [{\citenamefont {Kune\v{s}}(2014)}]{Kunes001}%
  \BibitemOpen
  \bibfield  {author} {\bibinfo {author} {\bibfnamefont {J.}~\bibnamefont
  {Kune\v{s}}},\ }\href {\doibase 10.1103/PhysRevB.90.235140} {\bibfield
  {journal} {\bibinfo  {journal} {Phys. Rev. B}\ }\textbf {\bibinfo {volume}
  {90}},\ \bibinfo {pages} {235140} (\bibinfo {year} {2014})}\BibitemShut
  {NoStop}%
\bibitem [{\citenamefont {Kune\v{s}}\ and\ \citenamefont
  {Augustinsk\'{y}}(2014{\natexlab{b}})}]{Kunes002}%
  \BibitemOpen
  \bibfield  {author} {\bibinfo {author} {\bibfnamefont {J.}~\bibnamefont
  {Kune\v{s}}}\ and\ \bibinfo {author} {\bibfnamefont {P.}~\bibnamefont
  {Augustinsk\'{y}}},\ }\href {\doibase 10.1103/PhysRevB.90.235112} {\bibfield
  {journal} {\bibinfo  {journal} {Phys. Rev. B}\ }\textbf {\bibinfo {volume}
  {90}},\ \bibinfo {pages} {235112} (\bibinfo {year}
  {2014}{\natexlab{b}})}\BibitemShut {NoStop}%
\bibitem [{\citenamefont {Kaneko}\ \emph {et~al.}(2012)\citenamefont {Kaneko},
  \citenamefont {Seki},\ and\ \citenamefont {Ohta}}]{Kaneko2012}%
  \BibitemOpen
  \bibfield  {author} {\bibinfo {author} {\bibfnamefont {T.}~\bibnamefont
  {Kaneko}}, \bibinfo {author} {\bibfnamefont {K.}~\bibnamefont {Seki}}, \ and\
  \bibinfo {author} {\bibfnamefont {Y.}~\bibnamefont {Ohta}},\ }\href {\doibase
  10.1103/PhysRevB.85.165135} {\bibfield  {journal} {\bibinfo  {journal} {Phys.
  Rev. B}\ }\textbf {\bibinfo {volume} {85}},\ \bibinfo {pages} {165135}
  (\bibinfo {year} {2012})}\BibitemShut {NoStop}%
\bibitem [{\citenamefont {Kaneko}\ and\ \citenamefont
  {Ohta}(2014)}]{Kaneko2014}%
  \BibitemOpen
  \bibfield  {author} {\bibinfo {author} {\bibfnamefont {T.}~\bibnamefont
  {Kaneko}}\ and\ \bibinfo {author} {\bibfnamefont {Y.}~\bibnamefont {Ohta}},\
  }\href {\doibase 10.1103/PhysRevB.90.245144} {\bibfield  {journal} {\bibinfo
  {journal} {Phys. Rev. B}\ }\textbf {\bibinfo {volume} {90}},\ \bibinfo
  {pages} {245144} (\bibinfo {year} {2014})}\BibitemShut {NoStop}%
\bibitem [{\citenamefont {Kaneko}\ \emph {et~al.}(2015)\citenamefont {Kaneko},
  \citenamefont {Zenker}, \citenamefont {Fehske},\ and\ \citenamefont
  {Ohta}}]{Kaneko2015}%
  \BibitemOpen
  \bibfield  {author} {\bibinfo {author} {\bibfnamefont {T.}~\bibnamefont
  {Kaneko}}, \bibinfo {author} {\bibfnamefont {B.}~\bibnamefont {Zenker}},
  \bibinfo {author} {\bibfnamefont {H.}~\bibnamefont {Fehske}}, \ and\ \bibinfo
  {author} {\bibfnamefont {Y.}~\bibnamefont {Ohta}},\ }\href {\doibase
  10.1103/PhysRevB.92.115106} {\bibfield  {journal} {\bibinfo  {journal} {Phys.
  Rev. B}\ }\textbf {\bibinfo {volume} {92}},\ \bibinfo {pages} {115106}
  (\bibinfo {year} {2015})}\BibitemShut {NoStop}%
\bibitem [{\citenamefont {Sotnikov}\ and\ \citenamefont
  {Kune\v{s}}()}]{Sotnikov}%
  \BibitemOpen
  \bibfield  {author} {\bibinfo {author} {\bibfnamefont {A.}~\bibnamefont
  {Sotnikov}}\ and\ \bibinfo {author} {\bibfnamefont {J.}~\bibnamefont
  {Kune\v{s}}},\ }\href@noop {} {\ }\Eprint
  {http://arxiv.org/abs/arXiv:1604.01997} {arXiv:1604.01997} \BibitemShut
  {NoStop}%
\bibitem [{\citenamefont {Murakami}\ \emph {et~al.}(1998)\citenamefont
  {Murakami}, \citenamefont {Hill}, \citenamefont {Gibbs}, \citenamefont
  {Blume}, \citenamefont {Koyama}, \citenamefont {Tanaka}, \citenamefont
  {Kawata}, \citenamefont {Arima}, \citenamefont {Tokura}, \citenamefont
  {Hirota},\ and\ \citenamefont {Endoh}}]{Murakami}%
  \BibitemOpen
  \bibfield  {author} {\bibinfo {author} {\bibfnamefont {Y.}~\bibnamefont
  {Murakami}}, \bibinfo {author} {\bibfnamefont {J.~P.}\ \bibnamefont {Hill}},
  \bibinfo {author} {\bibfnamefont {D.}~\bibnamefont {Gibbs}}, \bibinfo
  {author} {\bibfnamefont {M.}~\bibnamefont {Blume}}, \bibinfo {author}
  {\bibfnamefont {I.}~\bibnamefont {Koyama}}, \bibinfo {author} {\bibfnamefont
  {M.}~\bibnamefont {Tanaka}}, \bibinfo {author} {\bibfnamefont
  {H.}~\bibnamefont {Kawata}}, \bibinfo {author} {\bibfnamefont
  {T.}~\bibnamefont {Arima}}, \bibinfo {author} {\bibfnamefont
  {Y.}~\bibnamefont {Tokura}}, \bibinfo {author} {\bibfnamefont
  {K.}~\bibnamefont {Hirota}}, \ and\ \bibinfo {author} {\bibfnamefont
  {Y.}~\bibnamefont {Endoh}},\ }\href {\doibase 10.1103/PhysRevLett.81.582}
  {\bibfield  {journal} {\bibinfo  {journal} {Phys. Rev. Lett.}\ }\textbf
  {\bibinfo {volume} {81}},\ \bibinfo {pages} {582} (\bibinfo {year}
  {1998})}\BibitemShut {NoStop}%
\bibitem [{\citenamefont {McQueen}\ \emph {et~al.}(2008)\citenamefont
  {McQueen}, \citenamefont {Stephens}, \citenamefont {Huang}, \citenamefont
  {Klimczuk}, \citenamefont {Ronning},\ and\ \citenamefont {Cava}}]{Mcqueen}%
  \BibitemOpen
  \bibfield  {author} {\bibinfo {author} {\bibfnamefont {T.~M.}\ \bibnamefont
  {McQueen}}, \bibinfo {author} {\bibfnamefont {P.~W.}\ \bibnamefont
  {Stephens}}, \bibinfo {author} {\bibfnamefont {Q.}~\bibnamefont {Huang}},
  \bibinfo {author} {\bibfnamefont {T.}~\bibnamefont {Klimczuk}}, \bibinfo
  {author} {\bibfnamefont {F.}~\bibnamefont {Ronning}}, \ and\ \bibinfo
  {author} {\bibfnamefont {R.~J.}\ \bibnamefont {Cava}},\ }\href {\doibase
  10.1103/PhysRevLett.101.166402} {\bibfield  {journal} {\bibinfo  {journal}
  {Phys. Rev. Lett.}\ }\textbf {\bibinfo {volume} {101}},\ \bibinfo {pages}
  {166402} (\bibinfo {year} {2008})}\BibitemShut {NoStop}%
\bibitem [{\citenamefont {Vogt}\ \emph {et~al.}(2000)\citenamefont {Vogt},
  \citenamefont {Woodward}, \citenamefont {Karen}, \citenamefont {Hunter},
  \citenamefont {Henning},\ and\ \citenamefont {Moodenbaugh}}]{Vogt}%
  \BibitemOpen
  \bibfield  {author} {\bibinfo {author} {\bibfnamefont {T.}~\bibnamefont
  {Vogt}}, \bibinfo {author} {\bibfnamefont {P.~M.}\ \bibnamefont {Woodward}},
  \bibinfo {author} {\bibfnamefont {P.}~\bibnamefont {Karen}}, \bibinfo
  {author} {\bibfnamefont {B.~A.}\ \bibnamefont {Hunter}}, \bibinfo {author}
  {\bibfnamefont {P.}~\bibnamefont {Henning}}, \ and\ \bibinfo {author}
  {\bibfnamefont {A.~R.}\ \bibnamefont {Moodenbaugh}},\ }\href {\doibase
  10.1103/PhysRevLett.84.2969} {\bibfield  {journal} {\bibinfo  {journal}
  {Phys. Rev. Lett.}\ }\textbf {\bibinfo {volume} {84}},\ \bibinfo {pages}
  {2969} (\bibinfo {year} {2000})}\BibitemShut {NoStop}%
\bibitem [{\citenamefont {Nakao}\ \emph {et~al.}(2011)\citenamefont {Nakao},
  \citenamefont {Murata}, \citenamefont {Bizen}, \citenamefont {Murakami},
  \citenamefont {Ohoyama}, \citenamefont {Yamada}, \citenamefont {Ishiwata},
  \citenamefont {Kobayashi},\ and\ \citenamefont {Terasaki}}]{Nakao}%
  \BibitemOpen
  \bibfield  {author} {\bibinfo {author} {\bibfnamefont {H.}~\bibnamefont
  {Nakao}}, \bibinfo {author} {\bibfnamefont {T.}~\bibnamefont {Murata}},
  \bibinfo {author} {\bibfnamefont {D.}~\bibnamefont {Bizen}}, \bibinfo
  {author} {\bibfnamefont {Y.}~\bibnamefont {Murakami}}, \bibinfo {author}
  {\bibfnamefont {K.}~\bibnamefont {Ohoyama}}, \bibinfo {author} {\bibfnamefont
  {K.}~\bibnamefont {Yamada}}, \bibinfo {author} {\bibfnamefont
  {S.}~\bibnamefont {Ishiwata}}, \bibinfo {author} {\bibfnamefont
  {W.}~\bibnamefont {Kobayashi}}, \ and\ \bibinfo {author} {\bibfnamefont
  {I.}~\bibnamefont {Terasaki}},\ }\href {\doibase 10.1143/JPSJ.80.023711}
  {\bibfield  {journal} {\bibinfo  {journal} {J. Phys. Soc. Jpn.}\ }\textbf
  {\bibinfo {volume} {80}},\ \bibinfo {pages} {023711} (\bibinfo {year}
  {2011})}\BibitemShut {NoStop}%
\bibitem [{\citenamefont {Fujioka}\ \emph {et~al.}(2013)\citenamefont
  {Fujioka}, \citenamefont {Yamasaki}, \citenamefont {Nakao}, \citenamefont
  {Kumai}, \citenamefont {Murakami}, \citenamefont {Nakamura}, \citenamefont
  {Kawasaki},\ and\ \citenamefont {Tokura}}]{Fujioka}%
  \BibitemOpen
  \bibfield  {author} {\bibinfo {author} {\bibfnamefont {J.}~\bibnamefont
  {Fujioka}}, \bibinfo {author} {\bibfnamefont {Y.}~\bibnamefont {Yamasaki}},
  \bibinfo {author} {\bibfnamefont {H.}~\bibnamefont {Nakao}}, \bibinfo
  {author} {\bibfnamefont {R.}~\bibnamefont {Kumai}}, \bibinfo {author}
  {\bibfnamefont {Y.}~\bibnamefont {Murakami}}, \bibinfo {author}
  {\bibfnamefont {M.}~\bibnamefont {Nakamura}}, \bibinfo {author}
  {\bibfnamefont {M.}~\bibnamefont {Kawasaki}}, \ and\ \bibinfo {author}
  {\bibfnamefont {Y.}~\bibnamefont {Tokura}},\ }\href {\doibase
  10.1103/PhysRevB.92.195115} {\bibfield  {journal} {\bibinfo  {journal} {Phys.
  Rev. Lett.}\ }\textbf {\bibinfo {volume} {111}},\ \bibinfo {pages} {027206}
  (\bibinfo {year} {2013})}\BibitemShut {NoStop}%
\end{thebibliography}%
\end{document}